\documentclass[draft,jgrga]{agutex}

\usepackage{graphicx}

\setkeys{Gin}{draft=false}

\authorrunninghead{PULUPA ET AL.}

\titlerunninghead{STEREO FORESHOCK ELECTRONS}

\authoraddr{M. P. Pulupa, Space Sciences Laboratory, University of California, 7 Gauss Way, Berkeley, CA 94708, USA. (pulupa@ssl.berkeley.edu)}

\authoraddr{S. D. Bale, Department of Physics and Space Sciences Laboratory, University of California, 7 Gauss Way, Berkeley, CA 94708, USA. (bale@ssl.berkeley.edu)}

\authoraddr{A. Opitz, IRAP-Universit\'{e} de Toulouse, CNRS, 9 av. du colonel Roche, BP 44346, 31028 Toulouse Cedex 4, France (andrea.opitz@cesr.fr)}

\authoraddr{A. Fedorov, IRAP-Universit\'{e} de Toulouse, CNRS, 9 av. du colonel Roche, BP 44346, 31028 Toulouse Cedex 4, France (fedorov@cesr.fr)}

\authoraddr{R. P. Lin, Space Sciences Laboratory, University of California, 7 Gauss Way, Berkeley, CA 94708, USA. (rlin@ssl.berkeley.edu)}

\authoraddr{J.-A. Sauvaud, IRAP-Universit\'{e} de Toulouse, CNRS, 9 av. du colonel Roche, BP 44346, 31028 Toulouse Cedex 4, France (jean‐andre.sauvaud@cesr.fr)}

\begin{document}

\title{STEREO measurements of electron acceleration beyond fast Fermi at the bow shock}

%
%

\authors{M. P. Pulupa, \altaffilmark{1}, 
S. D. Bale \altaffilmark{1,2}, 
A. Opitz \altaffilmark{3},
A. Fedorov \altaffilmark{3},
R. P. Lin \altaffilmark{1,4}, 
J.-A. Sauvaud \altaffilmark{3}}

\altaffiltext{1}{Space Sciences Laboratory, University of California,
Berkeley, California, USA}
\altaffiltext{2}{Physics Department, University of California,
Berkeley, California, USA}
\altaffiltext{3}{IRAP (CNRS-UPS), University of Toulouse, Toulouse, France}
\altaffiltext{4}{School of Space Research, Kyung Hee University, Yongin, Korea}
%
%


\begin{abstract}
  Solar wind electrons are accelerated and reflected upstream by the terrestrial bow shock into a region known as the electron foreshock.  Previously observed electron spectra at low energies are consistent with a fast Fermi mechanism, based on the adiabatic conservation of the magnetic moment ($\mu$) of the accelerated electrons.  At higher energies, suprathermal power law tails are observed beyond the level predicted by fast Fermi.

  The SWEA and STE electron detectors on STEREO enable measurements of foreshock electrons with good energy resolution and sensitivity over the entire foreshock beam.  We investigate the electron acceleration mechanism by comparing observed STEREO electron spectra with predictions based on a Liouville mapping of upstream electrons through a shock encounter.  The foreshock electron beam extends up to several tens of keV, energies for which the Larmor radii of electrons is tens of km or greater.  These radii are comparable to the scale sizes of the shock, and $\mu$ conservation no longer applies.  We show that the observed enhancement in the foreshock beam beyond fast Fermi levels begins at energies where this assumption breaks down.

  We also demonstrate, using the Liouville mapping technique, that the strahl plays an important role in the formation of the bump on tail instability. We discuss this in the context of recent observations in the foreshock and in solar wind magnetic holes.
\end{abstract}

%
%

%

\begin{article}

%
%

\section{Introduction}\label{sec:introduction}

The region of interplanetary space which is upstream of the terrestrial bow shock and magnetically connected to the shock is known as the foreshock. Observations of upstream electron beams \citep{1979GeoRL...6..401A, 1983JGR....88...96F, 1989JGR....9410011G} near the leading edge of the foreshock demonstrated that incident solar wind electrons are accelerated by the quasiperpendicular bow shock back into the interplanetary medium.  These electron beams are frequently associated with electrostatic Langmuir oscillations. \citet{1979JGR....84.1369F} pointed out that the effect of velocity dispersion of accelerated electrons convecting downstream with the solar wind is to create bump on tail electron velocity distribution functions (eVDFs).  The bump on tail eVDFs are then unstable to the growth of Langmuir waves via Landau resonance.

The canonical picture of the quasiperpendicular electron acceleration process is the Fast Fermi model developed by \citet{1984AnGeo...2..449L} and \citet{1984JGR....89.8857W}, which treats the electron-shock encounter assuming conservation of the first adiabatic invariant ($\mu = mv_{\perp}^2/2B$) in the de Hoffmann-Teller (HT) frame.  In the HT frame, the upstream bulk velocity $\mathbf{u}_{\mathrm{u}}$ and magnetic field $\mathbf{B}_{\mathrm{u}}$ are parallel, and the electrons do not encounter a motional electric field.  The transformation velocity $v_{\mathrm{HT}}$ to the HT frame from any shock stationary frame lies in the plane of the shock and is given by

\begin{equation}
  \mathbf{v}_{\mathrm{HT}} =
  \frac{\hat{\mathbf{n}} \times 
  (\mathbf{u}_{\mathrm{u}} \times \mathbf{B}_{\mathrm{u}})}
  {\mathbf{B}_{\mathrm{u}} \cdot \hat{\mathbf{n}}}
  \label{eqn_vht}
\end{equation}

In this picture, the energization is a consequence of the boost to the HT frame, the mirroring, and the boost back to the solar wind frame. The accelerated electrons gain parallel velocity proportional to $2v_{\mathrm{HT}}$. Detailed expressions for the eVDFs upstream of the shock in the foreshock region have been presented by \citet{1987JGR....92.2315C} and \citet{1990JGR....95.4155F}.  A review of the physics of electron acceleration at the bow shock is presented in \citet{burgess:acceleration2007}.

The fast Fermi theory works well to describe relatively low energy electron acceleration.  Observed features of foreshock electrons such as prominent bump on tail eVDFs \citep{1984GeoRL..11..496F, 1996GeoRL..23.1235F} and energy dependent loss cone widths \citep{1996GeoRL..23.2203L} can be explained in the context of the theory and used to deduce information about the shock at the acceleration site.  Simulations of electron interactions with planar and curved shocks \citep{1989JGR....9415089K,1991JGR....96..143K} also agree well with the fast Fermi model.

However, for the more energetic electrons observed in the foreshock (for example, the $>16$ keV electrons in Figure~2 of \citet{1979GeoRL...6..401A}, and the power law tails observed by \citet{1989JGR....9410011G} and \citet{2006GeoRL..3324104O}), the assumptions used in the fast Fermi model are inapplicable.  Electrons with energies of tens of keV or higher have Larmor radii $r_{\mathrm{Le}}$ which are comparable to the ion Larmor radius $r_{\mathrm{Li}}$ and/or the ion inertial length $c/\omega_{\mathrm{pi}}$ in the solar wind at 1 AU. Since the scale size of the quasiperpendicular bow shock is set by these ion scales \citep{2003PhRvL..91z5004B}, the conditions for $\mu$ conservation are violated, and the presence of the observed suprathermal electrons must be explained by physics beyond single encounter fast Fermi acceleration.

This study presents STEREO observations of foreshock electron beams with good energy resolution over the entire energy range of the beams. Section \ref{sec:observations} describes the STEREO instrumentation, and Section \ref{sec:incident-electrons} describes the fitting procedure for the incident population of solar wind electrons. Section~\ref{sec:pred-foreshock-electrons} discusses the method of Liouville mapping the incident eVDF through the shock encounter.  The Liouville mapping predictions are compared with observations in Section~\ref{sec:obs-foreshock-electrons}, and the impact of the results on the theory of electron shock acceleration are discussed.  Section~\ref{sec:summary-conclusions} summarizes the conclusions of this study.

\section{STEREO Measurements}\label{sec:observations}

\subsection{Overview}\label{sec:obs-overview}

During the early part of the STEREO mission, both spacecraft spent significant time in the terrestrial electron foreshock. Upstream Langmuir waves \citep{2009JGRA..11412101M} and electron beams were observed over distances of several hundred $R_\mathrm{E}$ upstream of the shock. This paper focuses on one foreshock event, which occurred at approximately 0730 UT on 20 December 2006.  At this time, the spacecraft GSE coordinates were approximately $[77.9, -64.4, -10.8] R_E$.

Langmuir wave observations were made with the S/WAVES LFR instrument \citep{2008SSRv..136..487B}, which measures electric field fluctuations from 2.5 to 160 kHz, a range which includes the local plasma frequency.  Magnetic field measurements were made with the STEREO/IMPACT MAG instrument \citep{2008SSRv..136..117L, 2008SSRv..136..203A}.  Solar wind and foreshock electron observations were made with the STEREO/IMPACT electron detectors.  The instruments used were the electrostatic SWEA detector \citep{2008SSRv..136..227S}, which measures electrons from 1 eV to 2 keV with nearly $4\pi$ angular coverage, and the solid state detector STE \citep{2008SSRv..136..241L}, which measures electrons from 2 keV to 100 keV.  Together, these detectors span the energy range of the known populations of quiet time solar wind electrons \citep{1998SSRv...86...61L}.

\begin{figure}
\centering\includegraphics[width=26pc]{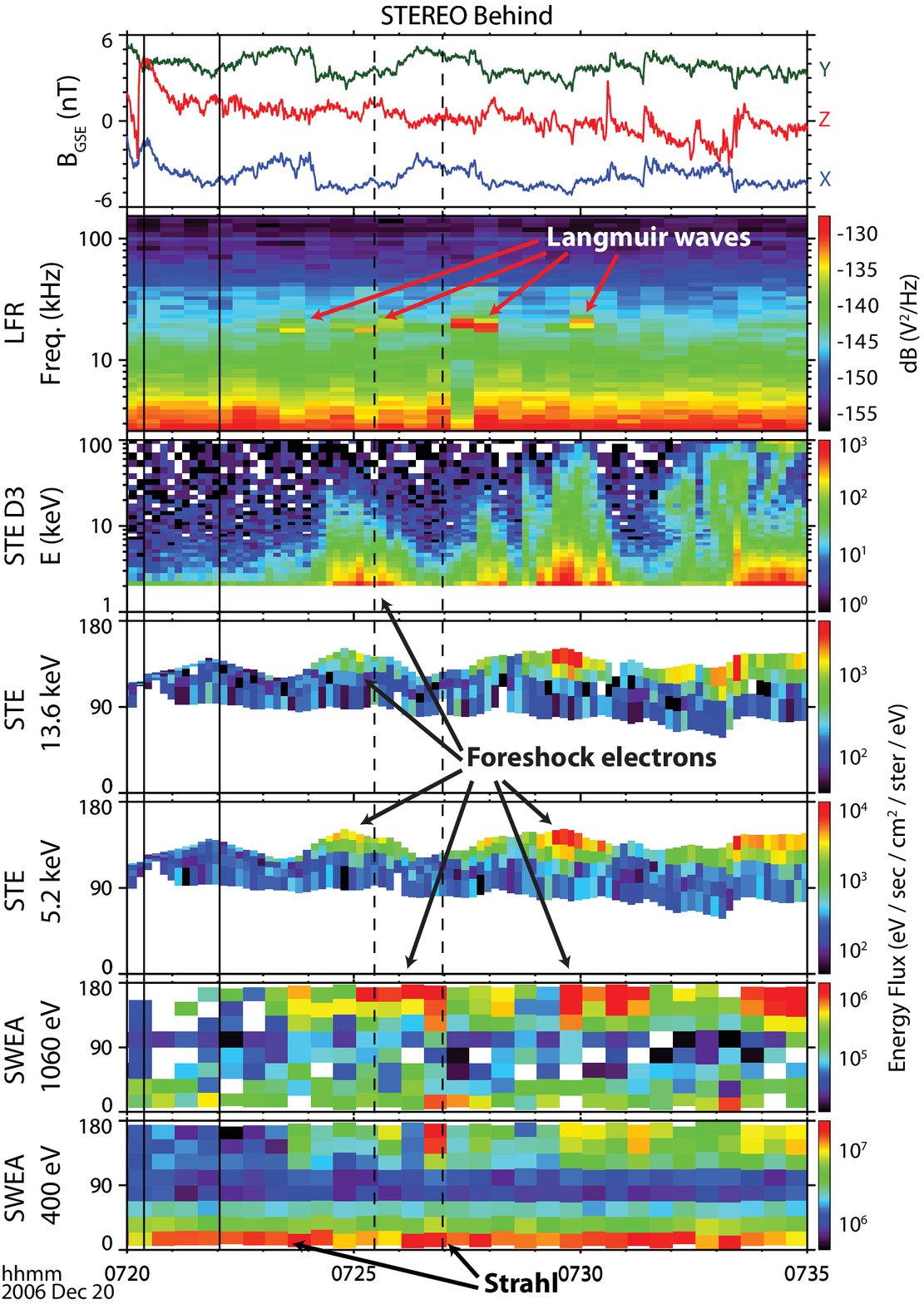}
  \caption{Upstream foreshock electron event observed by STEREO/Behind on 20 December 2006.  From top to bottom, the plotted quantities are: GSE magnetic field, electric field spectrogram from S/WAVES LFR, energy spectrogram from the STE D3 detector, STE PADs at 13.6 keV and 5.2 keV, and SWEA PADs at 1060 eV and 400 eV.  The foreshock electrons are visible in the STE spectrogram starting at around 7:25 UT.  The PADs for the electrons show enhancements in the antiparallel direction corresponding to foreshock electron beams.  The time interval delineated by the two vertical bars near 7:20 UT is used to calculate the incident eVDF.  The time intervals near 7:25 and 7:27 UT correspond to the two foreshock intervals analyzed in Section~\ref{sec:obs-foreshock-electrons}.}
\label{fig_tplot}
\end{figure}

An overview plot of the foreshock event is shown in Figure~\ref{fig_tplot}.  From top to bottom, Figure~\ref{fig_tplot} shows the magnetic field in GSE coordinates, the LFR electric field spectrogram, an energy spectrogram from the STE D3 detector, STE pitch angle distributions (PADs) at 13.6 keV and 5.2 keV, and SWEA PADs at 1060 eV and 400 eV.  For the entire event, the $\mathbf{x}_{\mathrm{GSE}}$ component of the magnetic field is negative, which implies that electrons with pitch angles close to $0^{\circ}$ are traveling towards the bow shock (i.e., incident to the shock) and those with pitch angles close to $180^{\circ}$ are traveling sunward (i.e., backstreaming from the shock).

Three time intervals are indicated in Figure~\ref{fig_tplot}.  During
the first interval (indicated by two solid vertical lines), from 0720 to 0722 UT, the spacecraft was in the undisturbed solar wind, unconnected to the bow shock.  The electron PADs for this time period show evidence of a beam-like strahl population, apparent as an enhancement in pitch angles close to $0^{\circ}$ in the 400 eV PAD.

The spacecraft then enters the electron foreshock.  Starting just before 0725 UT, foreshock electron beams appear as enhancements in pitch angles close to $180^{\circ}$.  The foreshock electron beams are visible in the SWEA and STE PADs.  Bursty beam-driven Langmuir wave emission at around 19 kHz is also apparent. The STE energy spectrogram shows that the foreshock electron beam evolves rapidly.  This evolution is an expected and previously observed consequence of the time-of-flight velocity dispersion in the foreshock \citep{1979JGR....84.1369F}.  



The second and third time intervals (indicated in Figure~\ref{fig_tplot} by vertical dashed lines) show observations at two different foreshock depths. During the first interval at 07:25, the spacecraft is close to the foreshock edge, and the foreshock beam energy spectrum reaches well into the STE energy range.  During the second interval at 07:27, the spacecraft is deeper in the foreshock and the foreshock beam is at lower energies.  These intervals will be examined in detail in Section~\ref{sec:obs-foreshock-electrons}.
 
\subsection{Interpretation of STE observations}\label{sec:ste-observations}

The STE instrument has two sensor units, each consisting of a set of four thin window silicon semiconductor detectors.  The two units are centered along the nominal Parker spiral in the upstream (STE-U) and downstream (STE-D) directions.  
\citep{2008SSRv..136..241L, 2010GeoRL..3708107W}.

Because the central look detection of each STE angular bin lies in the ecliptic, the pitch angle coverage of the STE instrument depends on the magnetic field observed at the spacecraft.  For example, if the field is entirely in the $\mathbf{z}_{\mathrm{GSE}}$ direction, each STE detector observes the same pitch angle ($90^{\circ}$).  If the field is in the $\mathbf{x}_{\mathrm{GSE}}-\mathbf{y}_{\mathrm{GSE}}$ plane, more pitch angle information can be recovered.  Evidence of this dependence is visible in the STE PAD plots in Figure~\ref{fig_tplot}, which demonstrate considerably wider range in pitch angle as the $\mathbf{z}_{\mathrm{GSE}}$ component of the field decreases.

When STEREO is in its normal orientation with the SECCHI imagers \citep{2008SSRv..136...67H} looking at the sun (which is the case for the entire mission, excepting some early orbital maneuvers and short, isolated magnetometer calibration intervals), multiply reflected solar optical and UV light contaminates the STE-U detectors.  For this reason, only STE-D data is presented in this paper.

Because the detectors are open, without curved path electrostatic deflectors or foil covers, STE measures ions, neutrals \citep{2010GeoRL..3708107W}, and X-rays \citep{2009ApJ...694L..79H} in addition to electrons.  For this reason, we must be careful in interpreting STE data in the foreshock region.  In order to be assured that an observed STE signal is indeed a foreshock electron beam, we require both (a) simultaneous electron measurements from SWEA and (b) associated Langmuir wave activity.  The STE observations themselves must also be consistent with electron beams.  An example of STE data which is inconsistent with foreshock electron beams is visible in Figure~\ref{fig_tplot} at 07:34 UT. Were the measured spectra at these times due solely to electrons, the features observable in the STE spectrogram at several tens of keV would be highly unstable, so the electron measurement at these times is likely contaminated by foreshock ions.

The two foreshock intervals highlighted in Figure \ref{fig_tplot} have STE spectra consistent with electrons, simultaneous SWEA observations, and associated S/WAVES Langmuir waves.  For the remainder of this paper the observations at these intervals are taken to be foreshock electron beams.


\section{Incident electrons}\label{sec:incident-electrons}

The solar wind electrons comprise several distinct populations; the populations considered in this paper are the core, the halo, the strahl, and the superhalo.  The thermal core population makes up most of the number density of the solar wind and can be modeled well using a bi-Maxwellian distribution.  The suprathermal halo population is better described by a bi-$\kappa$ distribution, which resembles a bi-Maxwellian at low energies and behaves like a power law distribution at high energies \citep{1997GeoRL..24.1151M}.  The strahl component of the solar wind is a beam-like feature propagating along the magnetic field in the antisunward direction.  A detailed description of the core, halo, and strahl components can be found in \citet{2009JGRA..11405104S}.


In addition to the above components, the quiet time solar wind electrons include a population known as the superhalo, which can be approximated using a power law from energies of several keV up to 100 keV \citep{1996GeoRL..23.1211L, lwang.2012.superhalo}.  The superhalo electrons have been shown to be nearly isotropic \citep{1998SSRv...86...61L}.

\begin{figure}
 \centering\includegraphics[width=23pc]{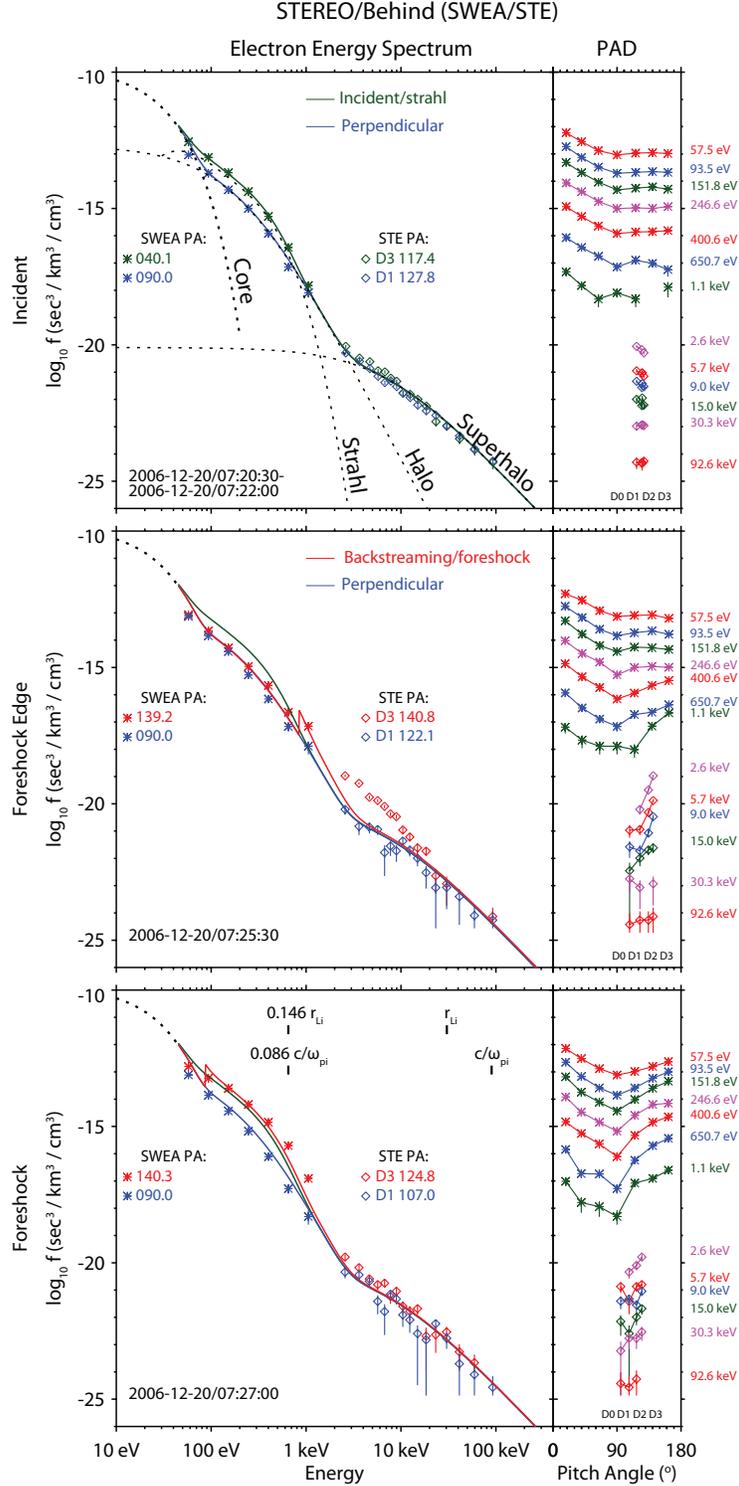}
  \caption{Energy spectra and PADs for the incident electrons (top row) and the foreshock electron beams at the foreshock edge (middle row) and deeper in the foreshock (bottom row).  SWEA pitch angle bins are marked by asterisks, and STE bins are marked by diamonds.  A full description of this figure can be found in Sections \ref{sec:incident-electrons} and \ref{sec:obs-foreshock-electrons} of the text.}
  \label{fig_espec_epad}
\end{figure}

The energy spectrum and the PAD of the incident solar wind electrons are shown in the top row of Figure~\ref{fig_espec_epad}.  Electrons from 57.5 eV to 1.1 keV are measured by the SWEA instrument and are marked with asterisks.  SWEA exhibits a degradation in low energy electron measurements due to instrumental charging effects \citep{2011SSRv..161...49F}, so the energies below 50~eV are not plotted in Figure~\ref{fig_espec_epad}. SWEA measurements above 50~eV are corrected by the transmission function described in \citet{2011SSRv..161...49F}.  The incident PAD shown in Figure \ref{fig_espec_epad} shows evidence of a strong strahl beam (the peak near zero pitch angle) and a relatively isotropic halo population (the relatively flat lines at pitch angles greater than $90^{\circ}$).

Electrons from 2 keV to 100 keV are measured by STE and are marked with diamonds.  During the time interval when the incident electrons were measured, the magnetic field contained a strong component in the $\mathbf{z}_{\mathrm{GSE}}$ direction, and the angular coverage of the STE instrument was limited to a narrow range of pitch angles.  For clarity, not all of the STE electron energy bins are shown.

Incident energy spectra at near-parallel (green) and near-perpendicular (blue) pitch angles are shown in the top left of Figure \ref{fig_espec_epad}, again plotted with asterisks for SWEA and diamonds for STE.  For SWEA, the blue perpendicular cut measurements are measured with bins with a pitch angle of $90^{\circ}$, while the green parallel cut measures electrons with pitch angle of $37^{\circ}$.  We do not use the pitch angle closest to $0^{\circ}$, as these electrons will likely lie inside the loss cone and, according to the fast Fermi theory, will not be accelerated by the shock.  The lack of pitch angle coverage from the STE instrument during this time interval limits the ability to choose near-parallel and near-perpendicular cuts.  This is mitigated by the fact that the quiet time superhalo is nearly isotropic \citep{1998SSRv...86...61L}.

The blue and green solid lines on the incident energy spectrum plot in Figure~\ref{fig_espec_epad} are the results of fitting the eVDF to a model including the core, the halo, the strahl, and the superhalo.  Each individual population is shown as a dotted line.  The blue (perpendicular) solid line is the sum of the core, halo, and superhalo populations.  The green (parallel) solid line is the sum of the blue line and the strahl population.

Due to the previously mentioned instrumental charging effects, SWEA measurements of the core population are limited.  To characterize the core, we instead use convected electron measurements from the SWE instrument on board the \emph{Wind} spacecraft \citet{1995SSRv...71...55O}, calculating the time lag according to the method described in \citet{2009SoPh..256..365O}.  The time-shifted core density $n_{\mathrm{c}}$ is $4.72/\mathrm{cm}^3$, and the core electron temperature $T_{\mathrm{c}}$ is $8.82$ eV.  This density corresponds to a local plasma frequency of $\sim 19.6$ kHz, which matches the frequency of the observed LFR Langmuir waves.


The halo component for this interval is well respresented by an isotropic $\kappa$ distribution.  The strahl component is modeled by subtracting the near-perpendicular SWEA cut from the near-parallel SWEA cut, then fitting the excess parallel observation as a function of energy using a spline fit.  The superhalo is fitted with a modified $\kappa$-type distribution which accounts for the relativistic behavior of the highest energy electrons \citep{2006PPCF...48..203X, 2010SoPh..267..153P}.  It is assumed that the quiet-time superhalo is isotropic.



The model eVDF matches the data well over the range of the halo, strahl, and superhalo. The parameters for the fitted eVDFs are summarized in Table \ref{tab_eparam}.  


\section{Predicted foreshock electron properties}\label{sec:pred-foreshock-electrons}

As stated above, the aim of this paper is to compare observed foreshock electron observations to the canonical Fast Fermi theory of electron acceleration.  In order to use this theory to predict the properties of a foreshock electron beam, information about (a) the shock parameters, (b) the interaction of the electrons with the shock, and (c) evolution of the beam is necessary.  In this section, we discuss the estimation of these quantities.

\subsection{Shock parameters}\label{sec:shock-parameters}

According to Fast Fermi theory, the acceleration of an electron by the bow shock depends on the loss cone angle $\alpha$, which is determined by the magnetic mirror ratio between the upstream magnetic field and the field in the shock front, the HT speed $v_{\mathrm{HT}}$, and the cross shock potential $\Phi$.  In order to determine the eVDF at an arbitrary point in the foreshock region, it is also necessary to know the depth of the spacecraft in the foreshock region, since the depth determines the cutoff electron speed $v_{\mathrm{c}}$ necessary for an electron to reach the spacecraft (electrons with $v < v_{\mathrm{c}}$ will be convected downstream by the motional electric field of the solar wind and will not be observed at the spacecraft \citep{1979GeoRL...6..401A, 1979JGR....84.1369F, 1987JGR....92.2315C}).

None of these parameters can be directly measured when the spacecraft is far from the shock.  On 20 December 2006, STEREO/B was approximately 100 $R_{\mathrm{E}}$ from of the Earth's bow shock, and several days had passed since its previous bow shock crossing.  Furthermore, the location of that previous crossing was not close to the quasiperpendicular connection point at which the electron acceleration takes place.  For these reasons, $\alpha$ and $\Phi$ cannot be accurately estimated using data from the previous STEREO shock crossing.  Previous papers dealing with electron acceleration at the bow shock (e.g., \citet{1979JGR....84.1369F, 1996GeoRL..23.2203L}) have used the straight field line approximation to determine $\theta_{\mathrm{bn}}$, which yields $v_{\mathrm{HT}}$, and depth in the foreshock, which yields $v_{c}$.  However, with a spacecraft more than several tens of $R_{E}$ upstream of the bow shock, magnetic field line wandering makes such an assumption invalid \citep{1996GeoRL..23..793Z}.  We must therefore estimate the relevant parameters using different methods.



\begin{figure}
  \centering\includegraphics[width=35pc]{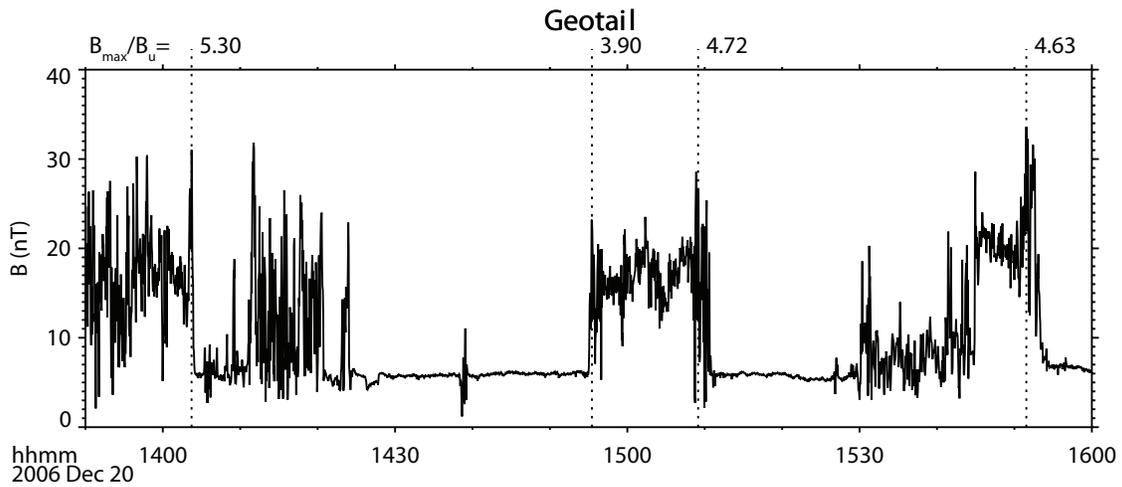}
  \caption{Several crossings of the quasiperpendicular bow shock as observed by Geotail on 20 December 2006.  Geotail crossed the shock near the electron acceleration site for the event observed by STEREO/B.  The crossings are indicated by dotted lines.  For each analyzed crossing, the ratio $B_{\mathrm{max}} / B_{\mathrm{u}}$ is shown.  A full description of this figure can be found in Section \ref{sec:shock-parameters}.}
  \label{fig_geotail}
\end{figure}

The ratio of the maximum field in the shock layer (including any overshoot) to the upstream field $B_{\mathrm{max}}/B_{\mathrm{u}}$ determines the opening angle $\alpha$ of the fast Fermi loss cone.  We use data from the Geotail spacecraft to estimate this parameter.  On 20 December 2006 at approximately 15:00 UT, Geotail observed several crossings of the quasiperpendicular bow shock at GSE coordinates of approximately $[2.7, -17.6, 3.8] R_{\mathrm{E}}$, i.e., in the neighborhood of the point where the STEREO/B spacecraft was connected to the shock some hours earlier.  Although this conjunction is only approximate in both time and space, the Geotail data nevertheless give a picture of the typical conditions of the quasiperpendicular shock during the relevant time interval.  Figure~\ref{fig_geotail} shows magnetic field data from the Geotail MGF instrument \citep{geotail_mgf} on 20 Dec 2006.  The quasiperpendicular shock crossings are seen as rapid changes in $|B|$.  The ratio $B_{\mathrm{max}}/B_{\mathrm{u}}$ is measured for several shock crossings and shown above the plot.  The ratios range from under 4 to slightly over 5, with an average $B_{\mathrm{max}}/B_{\mathrm{u}}$ of 4.64.  Using the relation $\sin^2 \alpha = B_{\mathrm{u}}/B_{\mathrm{max}}$ to compute the loss cone angle yields $\alpha = 27.7^{\circ}$.


In order to estimate $\Phi$, we use the result of an analytical model \citep{2002JGRA..107.1218K} for the cross shock potential which depends only on the upstream perpendicular electron temperature and the magnetic compression ratio.  

\begin{equation}
  e \Phi \simeq 2 k T_{\mathrm{e\perp u}} 
  \left(\frac{B_{\mathrm{d}}}{B_{\mathrm{u}}}-1\right)
  \label{eqn_phi}
\end{equation}

This model has been shown to be consistent with spacecraft measurements of electron temperature made during bow shock crossings \citep{2000JGR...10520957H}.  Using the convected \emph{Wind} electron measurements, the Geotail magnetic field, and Equation (\ref{eqn_phi}), we estimate the cross shock potential $\Phi \approx 72. \:\mathrm{V}$.



The $v_{\mathrm{HT}}$ parameter must be estimated from the foreshock electron data. Since electrons gain energy proportional to $2v_{\mathrm{HT}}$, the energy gain of the electrons can be used to estimate $v_{\mathrm{HT}}$, using the Liouville mapping technique described in the following section.  We use the relatively low energy SWEA electron energy bins to estimate $v_{\mathrm{HT}}$, by matching observed fluxes with predictions based on the upstream measurements. For the 07:25 UT interval at the foreshock edge, the estimated $v_{\mathrm{HT}}$ is $1500$ km/s.  For the 07:27 UT interval, the estimated $v_{\mathrm{HT}}$ is $500$ km/s.  Similarly, $v_{\mathrm{c}}$ can be estimated from the foreshock data by observing the lowest energy bin which shows evidence of foreshock electrons.  The estimated values of $v_{\mathrm{c}}$ for the 07:25 foreshock edge interval and the 07:27 interval are 13000 km/s and 4250 km/s, respectively.
 
The estimated Fast Fermi parameters $\alpha$, $\Phi$, $v_{\mathrm{HT}}$, and $v_{\mathrm{c}}$ are listed in Table \ref{tab_ffparam}.

\subsection{Liouville mapping}\label{sec:liouville-mapping}

Liouville's theorem states that, in the absence of collisions, the phase space density $f$ of a collection of particles remains constant along a trajectory defined by the (conservative and differentiable) forces which govern the system.  In the case of the solar wind electrons, this implies that $f_{\mathrm{i}}$ for the incident eVDF remains constant along trajectories determined by the fast Fermi theory, so long as that theory remains applicable.  The theorem allows for the prediction of foreshock eVDFs given the upstream observations and the shock parameters estimated above.

The Liouville theorem is the basis behind previous theoretical work predicting foreshock eVDFs \citep[e.g.][]{1987JGR....92.2315C, 1990JGR....95.4155F}.  It has also been used in experimental studies analyzing the evolution of the eVDF through bow shock crossings \citep{1989JGR....9410011G, 2001JGR...10615711H, 2007JGRA..11209212L}.  In the HT frame, electrons traveling toward the shock which satisfy the condition

\begin{equation}
\frac{1}{2}m\frac{v^{\prime2}_{\perp}}{B_{\mathrm{u}}} \ge 
\frac{1}{B_{\mathrm{d}}-B_{\mathrm{u}}}
\left(e\Phi' + \frac{1}{2}mv^{\prime2}_{\parallel}\right)
\label{eqn_refcrit}
\end{equation}

will be reflected by the shock, assuming conservation of the adiabatic invariant $\mu$ (see, e.g., \citet[Eq. 9]{2001JGR...10625041K}).  In the HT frame, these electrons reverse the sign of their parallel velocity, which adds a parallel velocity of $2v_{\mathrm{HT}}$.  A schematic diagram of fast Fermi acceleration which combines the illustrations of \citet{1984AnGeo...2..449L} and \citet{1984JGR....89.8857W} is shown in Figure 1 of \citet{2010JGRA..11504106P}.

In order to be observed at the spacecraft, the electrons must additionally meet the criterion that their parallel velocity is higher than the geometric cutoff velocity at the spacecraft.  The Liouville mapping of the incident eVDF $f_{\mathrm{i}}$ to the foreshock eVDF $f_{\mathrm{f}}$ can then be summarized in the following equation:

\begin{equation}
f_{\mathrm{f}} (v_{\parallel},v_{\perp}) =
\left\{%
\begin{array}{c@{\quad}l@{\:}l}
  f_{\mathrm{i}} (v_{\parallel},v_{\perp}) & 
  v_{\parallel}< & \mathrm{max}[v_{\mathrm{c}},v_{\mathrm{HT}}]
  \\
  & & \\
  f_{\mathrm{i}} (v_{\parallel},v_{\perp}) & 
  v_{\parallel}> & \mathrm{max}[v_{\mathrm{c}},v_{\mathrm{HT}}],
  \\
  & & \mathrm{in\:loss\:cone} \\ 
  f_{\mathrm{i}}(2 v_{\mathrm{HT} - }v_{\parallel},v_{\perp}) & 
  v_{\parallel}>& \mathrm{max}[v_{\mathrm{c}},v_{\mathrm{HT}}], 
  \\
  & & \mathrm{reflected} 
\end{array}%
\right.
\label{eqn_liouville}
\end{equation}

Using Equations~\ref{eqn_refcrit} and \ref{eqn_liouville}, we can directly calculate the predicted foreshock eVDFs given by Fast Fermi theory, and observe deviations from the measured STEREO data.  

\subsection{Quasilinear relaxation}\label{sec:quas-relax}

Foreshock electron beams can create a bump on tail feature which is unstable to the growth of Langmuir waves on very short time scales.  The foreshock beam is unstable when the reduced parallel distribution $f_{\parallel}(v_{\parallel})$ displays a positive slope in the $v_{\parallel}$ direction ($\partial f_{\parallel}/\partial v_{\parallel} > 0$).  In the quasilinear theory of beam relaxation \citep{1975AuJPh..28..731G, 1990SoPh..130..201M, 2000SoPh..194..345R}, positive slopes in the eVDF are mediated by Landau resonance, which generates Langmuir waves.  In short order, the region around the bump on tail evolves to a marginally stable ($\partial f_{\parallel}/\partial v_{\parallel} \approx 0$) plateaued state.

In this work, the method of \citet{2001JGR...10625041K} and \citet{2004JGRA..10902108K} is used in order to estimate the region over which the flattening of the beam will occur.  The procedure is as follows: Using the foreshock eVDF obtained from the Liouville mapping, the reduced distribution $f_{\parallel}$ is constructed.  The bump on tail is flattened by solving numerically for the points $v_{-}$ and $v_{+}$ which satisfy the conditions that $f_{\parallel}(v_{-}) = f_{\parallel}(v_{+})$ and the number of particles $\int_{v_{-}}^{v_{+}} f_{\parallel} d v_{\parallel}$ is conserved. The velocity of the beam $v_{\mathrm{b}}$ and its width $\Delta v_{\mathrm{b}}$ can then be calculated from $v_{-}$ and $v_{+}$.

\section{Foreshock electrons} \label{sec:obs-foreshock-electrons}

Energy spectra and PADs for the foreshock electrons are shown in the middle and bottom rows of Figure \ref{fig_espec_epad}, with the foreshock edge observations in the middle row and the deeper foreshock observations in the bottom row. As in the top row, energy spectra are to the left, PADs are to the right, SWEA data is plotted with asterisks, and STE data with diamonds.  The foreshock beam is evident in the foreshock PADs as an enhancement (compared to the solar wind PAD) for angles greater than $90^{\circ}$.  Near-perpendicular and near-antiparallel (backstreaming) cuts through the eVDF are plotted in the left panels with blue and red solid lines.  The solid lines plotted along with the electron spectra are the results of the Liouville mapping.

We analyze two intervals in order to emphasize different aspects of the foreshock electron beam.  Using the foreshock edge observations, we can isolate and examine the behavior of the high energy suprathermal foreshock beam electrons, lower energy electrons having been cut off from the beam by the foreshock geometry.  Using the deeper foreshock observations we can observe the lower energy foreshock beam.  These energies are useful for analyzing the effect of the strahl population and the limitations of the fast Fermi theory.

\subsection{Foreshock edge}\label{sec:foreshock-edge-obs}

The foreshock edge observations from 07:25 UT are shown in the middle row of Figure \ref{fig_espec_epad}.  For reference, the incident/strahl population line from the top row is also plotted on this spectrum as a green solid line.  A foreshock beam consisting of electrons which were simply reflected without energization would lie directly atop the green line.  Electron measurements showing evidence of energization therefore lie above the green line.  The electron beam in the foreshock edge energy spectrum is evident from the SWEA energy bin at 1.1 keV up to the 20 keV STE energy bin.  In terms of the Liouville mapping, this implies that the velocity cutoff is somewhere below 1.1 keV, and that $v_{\mathrm{HT}}$ can be determined by matching the observed beam at 1.1 keV to Liouville map predictions.  For the foreshock edge, we find $v_{c} \approx 13000$ km/s, and $v_{\mathrm{HT}} \approx 1500\:\mathrm{km/s}$.  However, it is clear that the Liouville predictions do not match the STE observations at higher energies.

This result is consistent with observations of \citet{1989JGR....9410011G}, who noted that a Liouville mapping predicting the downstream eVDF differed significantly from observations.  Additional processes beyond $\mu$-conserving reflection must apply for these suprathermal energies.  

\begin{figure}
 \centering\includegraphics[width=25pc]{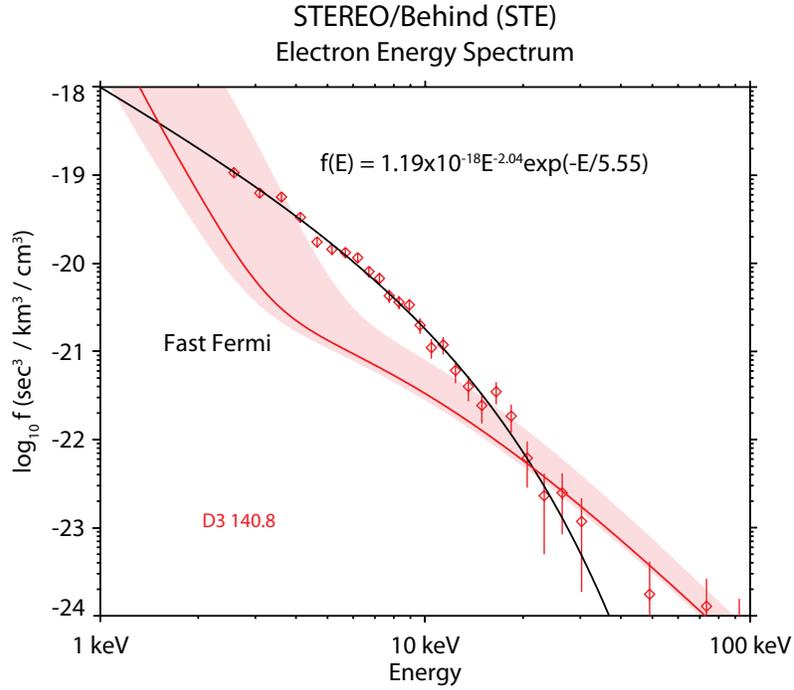}
 \caption{STE energy spectrum for the foreshock edge electron beam, demonstrating significant enhancement of accelerated electrons compared to the Fast Fermi theory.  The spectrum is well fit by a power law with an exponential roll off (black solid line).  The red line and the red shaded area show the range of predicted spectra calculated using Liouville's theorem and the fast Fermi acceleration mechanism.  A full description of this figure can be found in Section \ref{sec:foreshock-edge-obs}.}
 \label{fig_espec_ste}
\end{figure}

Figure \ref{fig_espec_ste} shows only the parallel STE data from the foreshock edge energy spectrum of Figure \ref{fig_espec_epad}.  The solid red line is the predicted spectrum using the fast Fermi model, as is also shown in Figure \ref{fig_espec_epad}.  The red shaded area in Figure \ref{fig_espec_ste} brackets the model predictions with extreme values of $v_{\mathrm{HT}}$.  Varying $v_{\mathrm{HT}}$ between $0$ and $10000\:\mathrm{km/s}$ yields the lower and upper boundaries of the shaded area, respectively.  It is clear that both the level and shape of the observed spectrum differs from the model predictions for the low and high values of $v_{\mathrm{HT}}$, emphasizing that the departure of the observations from fast Fermi is not simply a consequence of parameter choice.  The data are fit to an power law with an exponential roll-off, following \citet{2006GeoRL..3324104O}.  For this particular interval near the foreshock edge, we find a relatively energetic power law, with a power law index $\Gamma$ of $2.04$ and a roll-off energy $E_0$ of $5.5$ keV.  The power law index and rollover energy evolve as a function of time with variation in the electron beam, and the relationship between $\Gamma$ and $E_0$ may be the subject of a future study.


\subsection{Mechanisms for high energy electrons} \label{sec:mech-high-energy}

Previously, \citet{1989JGR....9410011G} noted that observed electron fluxes could not be adequately explained by the fast Fermi mechanism, and suggested that the suprathermal upstream electrons could be leaked from a downstream population which had been accelerated by the shock.  For the mechanism for generation of the observed enhancements in suprathermal electrons, \citet{1989JGR....9410011G} proposed shock drift acceleration.  Shock drift theory has been shown to be equivalent to the fast Fermi mechanism \citep{1989JGR....9415367K} for single encounters, but also can be naturally extended to study multiple encounters between the electron and the shock \citep{2001PASA...18..361B}.  Shock drift theory \citep{2001JGR...106.1859V} has been used to study the parametric dependence of upstream electrons on shock parameters, and to compare observed and predicted spectra.  The results are qualitatively similar but suggest that additional processes are important, especially for the high energy (above 10 keV) electrons.

More recently, several alternative ideas have been proposed, which explain the enhanced suprathermal electrons as a consequence of some type of process which can create pitch angle scattering in an incident electron \citep{burgess:acceleration2007}.  The scattered trajectory enables the electron to make multiple encounters with the shock and be accelerated to high energies.  We briefly describe several proposed processes in the paragraphs below.

Recent simulation work has also considered the effect of non-ideal structure on shock fronts on suprathermal electron acceleration.  Using hybrid simulations and an injection profile of 100 eV electrons, \citet{2006ApJ...653..316B} investigated several effects of shock front ripples, finding a considerable increase of the electron maximum energy and electron acceleration from a broader range of $\theta_{\mathrm{bn}}$ than was the case for planar shocks.  The difference in the interaction between a rippled and a planar shock lies in the ability of electrons to encounter the shock multiple times, due to the connection of a magnetic field line to multiple points on the surface of a rippled shock.  Evidence of such multiply connected electron acceleration sites (although for larger length scales than studied by \citet{2006ApJ...653..316B}) has been observed at CME-driven interplanetary shocks and is related to the generation of Type II radio bursts \citep{1999GeoRL..26.1573B, 2008ApJ...676.1330P}.

Previous Geotail observations of energetic electrons show a connection between energetic electrons and upstream whistler waves \citep{2006GeoRL..3324104O}, showing a dependence of foreshock electron power law index on the whistler critical Mach number.  This dependence has been explained as accelerated foreshock electron beams generating wave activity which can scatter electrons back and forth across the shock in a mechanism akin to the diffusive shock acceleration observed in the ion foreshock \citep{2010PhRvL.104r1102A}.  In a single event analysis using data from a Geotail shock crossing, a gradual (as opposed to spiky) temporal profile of accelerated electrons combined with observed upstream whistlers has been observed \citep{2009EP&S...61..603O}.

\citet{2010ApJ...715..406G} investigated the effects of upstream turbulence, also using hybrid simulations.  Using test particle trajectories, \citet{2010ApJ...715..406G} directly demonstrated the effect of multiple shock crossings on electron energization. \citet{2010ApJ...715..406G} also found that the level of upstream turbulence affected the maximum attainable energy for an injection profile of 100 eV electrons, as higher levels of turbulence increased field line wandering and thus the effect of multiple shock encounters.

\citet{2001JGR...10612975S} and \citet{2010JGRA..11509104S} have used two dimensional particle simulation to directly simulate electron acceleration at the terrestrial bow shock.  By tracing test particle orbits, several different types of accelerated electrons are found, including a population of electrons consistent with Fast Fermi and populations which are temporarily trapped in the front or escaped downstream.  The latter populations make up the higher-energy portion of the foreshock beam observed in the simulations, and possibly correspond to the high energy power law tails observed in the hybrid codes and in observations.

Recent work on quasiperpendicular shocks shows evidence of nonstationarity driven by reflected ions on the time scale of the ion gyroperiod causing the shock front to self-reform dynamically \citep{2002PhPl....9.1192K, 2007GeoRL..3405107L, 2009JGRA..11403217L}. \citet{2002JGRA..107.1037L} have suggested nonstationarity as an explanation for the bursty nature of upstream electron events.  The nonstationarity on the shock front, which in simulation results generates a highly non-uniform shock surface, could provide a mechanism for the multiple encounters which can generate high energy electrons.

In future work, observations such as those presented in this paper may be of use in discriminating between the various mechanisms described above.  With the STE detector, parameters of the suprathermal electron spectrum such as $\Gamma$ and $E_0$ can be studied with good energy resolution up to the highest beam energies.  Systematic variation of these parameters with solar wind and shock parameters may vary differently among the proposed mechanisms described in this section, and comparison with observations may therefore offer insight into the physics of electron acceleration.

\subsection{Limit of Fast Fermi}\label{sec:non-adiab-behav}

The foreshock edge beam examined above exhibited a high velocity cutoff.  As a consequence, the lower energy bins were not observed in the foreshock beam.  Deeper in the foreshock, the foreshock electron beam extends to lower energies.  For the observations from 07:27 UT, shown in the bottom row of Figure \ref{fig_espec_epad}, the electron beam is apparent in the energy spectrum from below 100 eV and up to the first several STE energies at several keV.  Using the same method of estimating velocity parameters as used in the previous section, we find $v_{c} \approx 4250$ km/s, and $v_{\mathrm{HT}} \approx 500 \:\mathrm{km/s}$ for the deeper foreshock measurements.

Using the parameters from Table \ref{tab_ffparam}, we can again compare the fast Fermi predictions with the observed data.  The Liouville-mapped data (red line) fits the observations (red asterisks) well for the relatively low energy SWEA bins from 151 to 400 eV, with an energy beam which is slightly enhanced compared to the incident strahl population from the top row (green line).  This enhancement is consistent with fast Fermi acceleration.  However, an additional enhancement in the foreshock electron beam beyond anything consistent with Fast Fermi theory begins in the 650 eV SWEA energy bin.

For typical conditions at the terrestrial bow shock, the combined SWEA and STE energy range covers the transition between adiabatic and non-adiabatic electrons, as the electron gyroradii for electrons with energies of tens of eV are generally much smaller than the shock scale (defined by the ion inertial length or ion gyroradius \citep{2003PhRvL..91z5004B})., while the electron gyroradii at tens of keV are comparable to or greater than these lengths.  Using convected Wind measurements from the SWE Faraday cup instrument, it is possible to estimate the ion Larmor radius and the ion inertial length.  For this event, $r_{\mathrm{Li}}=74\;\mathrm{km}$ and $c/\omega_{\mathrm{pi}}=123\;\mathrm{km}$.  Setting the electron Larmor radius to be equivalent to these length scales requires electron energies of $87\;\mathrm{keV}$ and $32\;\mathrm{keV}$, respectively. Similarly, an electron with an energy of $650\;\mathrm{eV}$ has a gyroradius equivalent to $0.086\;r_{\mathrm{Li}}$ or $0.143\;c/\omega_{\mathrm{pi}}$. Since these ion scale lengths define the scale size of the shock, this confirms that the electron foreshock beam begins to diverge from Fast Fermi at the expected point: when $r_{\mathrm{Le}}$ is no longer much smaller than the scale size of the shock ramp and hence $\mu$ is not conserved. 

We note that \citet{1989JGR....9410011G} used a Liouville technique to map the upstream eVDF through the shock and compare to the observed downstream eVDF.  In Figure 11b of \citet{1989JGR....9410011G}, the comparison between the observed and Liouville-mapped eVDFs shows the same deviation from single encounter fast Fermi theory, with the observed eVDF substantially enhanced beginning at energies of several hundred eV.

\subsection{Role of the strahl}\label{sec:role-strahl}




Since the cross shock potential at the terrestrial bow shock is typically of the order of the core electron temperature or higher \citep{2000JGR...10520957H, 2001JGR...10615711H, 2002JGRA..107.1218K, 2007JGRA..11209212L}, the foreshock electron beam is constituted primarily of electrons from the suprathermal solar wind populations.  Simulations of electron acceleration and the resulting plasma wave activity \citep{1984JGR....89.8857W, 1989JGR....9415089K, 1991JGR....96..143K, 2004JGRA..10902108K, 2009JGRA..11412101M} model the upstream eVDF as a sum of core and halo populations, where the halo population is represented as either a hotter Maxwellian population or as a $\kappa$ distribution.

The effect of the geometrical velocity cutoff is to separate the foreshock eVDF into incident solar wind electrons $v_{\parallel} < v_{\mathrm{c}}$ and reflected electrons $v_{\parallel} > v_{\mathrm{c}}$ (See Equation \ref{eqn_liouville}).  In the case of a backstreaming beam consisting of reflected halo electrons, a significant bump on tail will form only if the incident electrons are strongly accelerated by a high $v_{\mathrm{HT}}$ at the acceleration point---that is, only in a narrow region where $\theta_{\mathrm{bn}}$ is very close to $90^{\circ}$.

In the case of a backstreaming beam where the incident eVDF has a strong strahl component, however, the bump on tail can be more easily formed.  If the ratio of strahl density to halo density $n_{\mathrm{s}}/n_{\mathrm{h}}$ is high, even a small energization will result in a significant bump when combined with the effects of the geometric cutoff---the geometric cutoff will produce an eVDF which consists of the incident halo below $v_{\mathrm{c}}$ but the reflected strahl above $v_{\mathrm{c}}$.  A wider range of $\theta_{\mathrm{bn}}$ connection points will be able to produce bump on tail eVDFs with a strong strahl component present. 

\begin{figure}
 \centering\includegraphics[width=20pc]{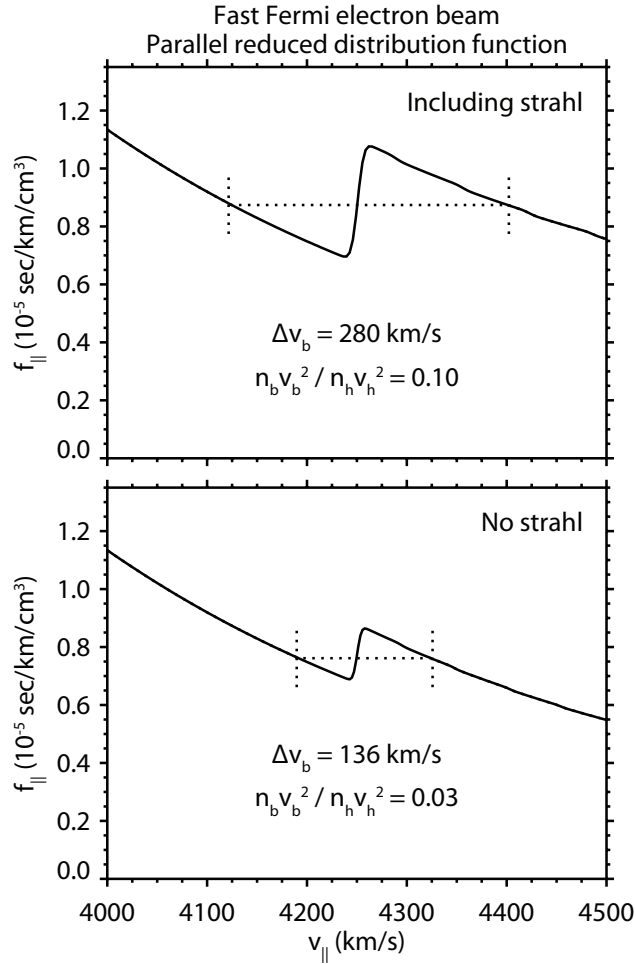}
 \caption{Reduced parallel distribution  $f_{\parallel}$ from the Liouville mapping.  The solid line in the upper plot shows $f_{\parallel}$ calculated using the parameters described in Sections~\ref{sec:incident-electrons} and \ref{sec:shock-parameters}.  The large bump on tail is a result of the strong strahl population in the incident eVDF.  The horizontal dotted line represents the relaxed plateau distribution constructed by eliminating the bump on tail while conserving particle number, as described in Section \ref{sec:quas-relax}.  The lower plot shows the results of the same procedure with the strahl removed from the incident eVDF, with the result that the bump is significantly smaller.  A full description of this figure can be found in Section \ref{sec:role-strahl}.}
\label{fig_qlrv}
\end{figure}

This effect is illustrated in Figure \ref{fig_qlrv}.  Using the eVDF parameters from Table \ref{tab_eparam}, the fast Fermi parameters for the deep foreshock from Table \ref{tab_ffparam}, and Equation \ref{eqn_liouville}, the reduced distribution $f_{\parallel}$ computed for two cases: one where the strahl population is included in the incident eVDF, and one where it is removed.  In each case, a bump on the tail of $f_{\parallel}$ is created by the fast Fermi process.  In the case where the strahl beam is included in the eVDF, the bump is higher and the beam is wider.  To quantify the difference in the bumps, the width of the beam $\Delta v_{\mathrm{b}}$ is calculated as described in Section~\ref{sec:quas-relax}.  A rough estimate for the energy in the electron beam is also calculated, by comparing $n_{\mathrm{b}} v_{\mathrm{b}}^2$ normalized to $n_{\mathrm{h}} v_{\mathrm{h}}^2$.  The beam from the eVDF including the strahl population is more than twice as wide in parallel $v$ and has more than three times the beam energy as does the beam from the eVDF without the strahl (see Figure \ref{fig_qlrv}).

The geometry of quasiperpendicular connection to the bow shock divides the electron foreshock into two separate wings, and that the strahl population is incident on the shock in only one wing. Given the discussion above, the presence of the reflected strahl in one foreshock wing will lead to an asymmetry in foreshock wave activity.  Such asymmetries have been noted, and correctly explained as a result of strahl electrons, using statistical studies of wave activity in the foreshock of Venus \citep{1993GeoRL..20.2801C, 1998JGR...10311985C}. Statistical analysis of electron and wave measurements has shown that the asymmetry also exists in the terrestrial foreshock \citep{2011GeoRL..3814105P}.  Recent results also connect observations of Langmuir waves in solar wind magnetic holes to the presence of a strahl population \citep{2010JGRA..11512113B}.  \citet{2010JGRA..11512113B} note that the suprathermal nature of the strahl population allow it to enter the magnetic hole and provide a source population for electron beams.  Similarly, the suprathermal nature of the strahl places the strahl electrons above the threshold condition for reflection determined by the cross shock potential $\Phi$.  In both the foreshock and in magnetic holes, the strahl adds asymmetry to the eVDF.  When the asymmetric eVDF interacts with a magnetic mirror-like structure such as a shock or the edge of a magnetic hole, the asymmetry leads to an enhancement in the bump on tail feature which is responsible for the generation of Langmuir waves.  

It is also possible that the microphysics of the generated Langmuir waves may be affected by the presence of the strahl.  Recent observations of type III-related Langmuir waves \citep{2011GeoRL..3813101M} have pointed to a relationship between electron beam speed and wave polarization.  By supplying a relatively high-energy seed population to the acceleration process, the strahl may indirectly affect wave polarization.

\section{Summary and Conclusions}\label{sec:summary-conclusions}

We have presented observations of foreshock electron beams using the STEREO spacecraft, combining the SWEA and STE electron instruments to make observations which simultaneously cover the entire range of foreshock electrons with good energy resolution and sensitivity.  Analysis of the beam electrons leads to the following conclusions.

Our results are consistent with the canonical Fast Fermi picture of foreshock electron acceleration for low energy electrons. A significant enhancement of electron acceleration begins at energies where the electron gyroradius is a significant fraction ($\sim 0.1$) of the ion gyroradius and/or ion inertial length.  The high energy tail of the foreshock electron beam is well fitted by a power law with an exponential roll-off, and extends up to several tens of keV before falling below the level of the background superhalo electron population.  As discussed in Section~\ref{sec:mech-high-energy}, a number of possible mechanisms have been shown to produce suprathermal electrons in this energy range.
 
Presence of the strahl is a significant factor in producing the bump on tail eVDFs which generate Langmuir waves in the electron foreshock. In particular, in upstream regions where the cutoff velocity $v_{\mathrm{c}}$ lies in the halo-strahl energy range, the asymmetry generated by a strong strahl implies that a large $v_{\mathrm{HT}}$ is not necessary to produce a significant bump.

The experimental results presented here may be useful for differentiating among the several physical mechanisms of the shock acceleration of electrons.  We suggest that simulation and modeling work in the future should include incident electrons featuring realistic strahl and superhalo populations in addition to halo electrons.


%

\begin{acknowledgments}
We are grateful to the STEREO IMPACT, MAG, and S/WAVES instrument teams.  We would like to specifically acknowledge gratitude to L. Wang for assistance with the STE data and to R. Gomez-Herrero for assistance with STEREO SEPT data.  We also thank I. Shinohara for access to the Geotail MGF data, and B. Maruca and M. Stevens for SWE Faraday cup data.  M. Pulupa acknowledges useful discussions with D. Burgess. This work was supported by NASA grant NNX10AM71G, and also supported in part by NASA grants NAS5-03131 and NNX09AG33G.  R. Lin was also supported in part by the WCU grant (R31-10016) funded by the Korean Ministry of Education, Science, and Technology. 
\end{acknowledgments}

%
%
%
%
%
%
%
%
%
%




%
%

\end{article}

\clearpage

\clearpage

\begin{table}     
  \centering
  \begin{tabular}{cc}
    $n_{\mathrm{c}}$ & 4.72 cm$^{-3}$ \\
    $T_{\mathrm{c}}$ & 8.82 eV \\
    $n_{\mathrm{h}}$ & 0.065 cm$^{-3}$ \\
    $T_{\mathrm{h}}$ & 48.6 eV \\
    $\kappa_{\mathrm{h}}$ & 5.9 \\
    $n_{\mathrm{sh}}$ & 0.731 $\times 10^{-6}$ cm$^{-3}$ \\
    $T_{\mathrm{sh}}$ & 2.50 keV\\
    $\kappa_{\mathrm{sh}}$ & 2.7
  \end{tabular}
  \caption{Measured parameters for the incident eVDF. The field-aligned strahl component is fit empirically using a spline fit and not to a parameterized model.}
\label{tab_eparam}
\end{table}

\begin{table}
  \centering
  \begin{tabular}{cc}
    $\alpha$ & $27.7^{\circ}$ \\
    $\Phi$ & $72. \:\mathrm{eV}$ \\
    $v_{\mathrm{HT}}$ (foreshock edge) & $1500 \:\mathrm{km/s}$ \\
    $v_{\mathrm{HT}}$ (deeper foreshock) & $500 \:\mathrm{km/s}$ \\
    $v_{\mathrm{c}}$ (foreshock edge) & $13000 \:\mathrm{km/s}$ \\
    $v_{\mathrm{c}}$ (deeper foreshock) & $4250 \:\mathrm{km/s}$
  \end{tabular}
 \caption{Parameters used to calculate the Liouville mapping of electrons through the fast Fermi shock encounter.  These parameters are estimated using upstream plasma measurements and Geotail shock crossings.  See Section \ref{sec:shock-parameters} for a complete description of parameter estimation.}
\label{tab_ffparam}
\end{table}

\end{document}